\documentclass[reprint]{JASA}
\usepackage{amsmath}
\usepackage{multirow}
\usepackage[caption=false]{subfig}
\usepackage{graphicx}%
\newsavebox{\measurebox}
\usepackage{array}

\usepackage{lipsum,capt-of,graphicx}
\usepackage[margin=1in]{geometry}
\usepackage{nicefrac, xfrac}
\begin{document}


\title[JASA Article]{ChildAugment: Data Augmentation Methods for Zero-Resource Children's Speaker Verification}







\author{Vishwanath Pratap Singh}
\affiliation{School of Computing, University of Eastern Finland, Joensuu, 80130, Finland}
\author{Md Sahidullah}
\affiliation{Institute for Advancing Intelligence, TCG CREST, Kolkata, West Bengal 700091, India and Academy of Scientific and Innovative Research (AcSIR), Ghaziabad- 201002, India}
\affiliation{Institute for Advancing Intelligence, TCG CREST, Kolkata, West Bengal 700091, India and Academy of Scientific and Innovative Research (AcSIR), Ghaziabad- 201002, India}
\author{Tomi Kinnunen}
\affiliation{School of Computing, University of Eastern Finland, Joensuu, 80130, Finland}
\preprint{Author, JASA}	

\begin{abstract}
The accuracy of modern automatic speaker verification (ASV) systems, when trained exclusively on adult data, drops substantially when applied to children's speech. The scarcity of children's speech corpora hinders fine-tuning ASV systems for children's speech. Hence, there is a timely need to explore more effective ways of reusing adults' speech data. One promising approach is to align vocal-tract parameters between adults and children through children-specific data augmentation, referred here to as \emph{ChildAugment}. Specifically, we modify the formant frequencies and formant bandwidths of adult speech to emulate children's speech. The modified spectra are used to train ECAPA-TDNN (emphasized channel attention, propagation, and aggregation in time-delay neural network) recognizer for children. We compare \emph{ChildAugment} against various state-of-the-art data augmentation techniques for children's ASV. We also extensively compare different scoring methods, including cosine scoring, PLDA (probabilistic linear discriminant analysis), and NPLDA (neural PLDA). We also propose a low-complexity weighted cosine score for extremely low-resource children ASV. Our findings on the CSLU kids corpus indicate that \emph{ChildAugment} holds promise as a simple, acoustics-motivated approach, for improving state-of-the-art deep learning based ASV for children. We achieve up to 12.45\% (boys) and 11.96\% (girls) relative improvement over the baseline.  {For reproducibility, we provide the evaluation protocols and codes \href{https://github.com/vpspeech/ChildAugment}{here}.}

\end{abstract}

\maketitle

\section{\label{sec:1} Introduction}

Children are often referred to as \emph{digi-native} because they have grown up in an era where digital technology is ubiquitous and easily accessible~\cite{diginative}. As a result, they have become a major group of daily smartphone, tablet, and digital device users \cite{oecd}. This increased digital proficiency sets important security and privacy requirements for the digital devices targeted at children, a group known to be particularly vulnerable to any harmful online activities. 

The current measures intended to prevent children from accessing harmful content, such as traditional password-based systems (which may be challenging for young children who cannot yet write/type), parental controls (which cannot block every potential risk), and self-declaration of age (which is easily circumvented), prove to be insufficient. In light of these limitations, speech technology emerges as a promising solution. Specifically, \emph{automatic speaker verification} (ASV) \cite{bai2021speaker} technology for verifying the user's identity based on his or her voice could be one of the solutions for preventing children from engaging in harmful online activities. {Besides strengthening security, ASV holds immense promise in personalizing children's experiences through seamless, user-friendly interaction with technology. Therein, ASV could enhance children's engagement in online education, enrich computer-assisted learning, and elevate the interactivity of speech-enabled toys and games. While our aim is to analyze and enhance algorithms for children ASV, it is clear that any practical deployment in this sensitive domain requires a careful evaluation of the technology's utility against privacy and ethical considerations, such as unauthorized user profiling.} Despite the wide range of potential applications of ASV in children's speech, research in children's ASV remains limited \cite{sembed,safavi}. 

{Modern deep learning based ASV systems \cite{xvector} generally involves three phases:} (1) training of a speaker embedding extractor to obtain a fixed-length feature vector that captures the speaker characteristics; 
(2) enrollment, where a reference model is generated after registering a speaker's voice; and (3) verification, which determines if a voice in an unlabeled test utterance matches the hypothesized reference voice. While modern ASV systems are relatively robust against a wide range of variability across training, enrollment, and test data, they are also highly sensitive to mismatches in the acoustic data domain across the training, testing, and verification phases. While the widely-studied factors responsible for degraded performance in ASV include mismatched channels, acoustic environments, or codecs, our main interest is on the relatively less studied issue of suppressing mismatch between adult and children speakers. 

Even if ASV technology has steadily improved for adult speech in recent years due to the advancements in deep learning \cite{bai2021speaker}, the accuracy degrades remarkably on children's voices. In prior studies \cite{shah, safavi}, a 40-45\% relative increase in \emph{equal error rate} (EER) is reported when embeddings from speaker embedding extractor trained on adult's speech is used in enrolling and verifying children's speech. This decline can be attributed to the mismatch between the training and evaluation conditions of the speaker embedding extractor. 
One of the primary reasons for this mismatch is the considerable disparity in vocal tract characteristics between adults and children, resulting from underdeveloped vocal tract and pronunciation skills, particularly in the age range of 3 to 14 years~\cite{ref5, jasa1}. 

The challenges of developing robust ASV systems for this age group are further convoluted by the limited supply of public corpora for children's speech~\cite{sembed} for training deep learning based children-specific speaker embedding extractor. {Some notable children-specific datasets include (i) The CSLU Kids Corpus \cite{cslu}, which contains both scripted and spontaneous recordings from over 1100 speakers, although it offers only a single session per speaker. (ii) PF-STAR \cite{pfstar}, containing around 15 hours of speech from 158 children aged 4 to 14 years. (iii) \emph{My Science Tutor (MyST)} \cite{myst}, which comprises 470 hours of English speech from 1371 students in grades 3-5 engaged in conversations. Even if a number of children corpora are readily available, they exhibit limitations in terms of the number of speakers, session variation, age diversity, and the total number of utterances, when compared to adult speech corpora such as VoxCeleb2 \cite{vox2}.}


\subsection{Challenges and Current Solutions for Children ASV}

In deep learning based \emph{automatic speech recognition} (ASR) and ASV tasks, there are various techniques for addressing domain mismatch, ranging from transfer learning~\cite{tlr} and feature normalization~\cite{pncasr,formantChild,f0modCASR} to various \emph{data augmentation methods}~\cite{aug1}. {Among these alternatives, the last one has been effective in improving children's ASR \cite{me2022,VTLNchildren,LPCAugment,kathania_aug} as it does not require children's speech and can be easily integrated with existing training pipelines.}
Despite its effectiveness in improving children's ASR \cite{kathania_aug,shah,me2022}, only a handful of studies have addressed children-specific data augmentations in ASV \cite{shahnawazuddin2020voice,safavi}. 
Moreover, none of the prior studies have explored the details of data augmentation, such as the impact of the proportion of original to augmented data and the proportion of different augmentations within the augmented data on ASV systems \cite{sb}. Additionally, the goals of ASR and ASV systems are fundamentally different.
Hence, a careful investigation is needed before applying the ASR-specific data augmentation to ASV tasks.

\begin{figure}[t]
\includegraphics[width=3in]{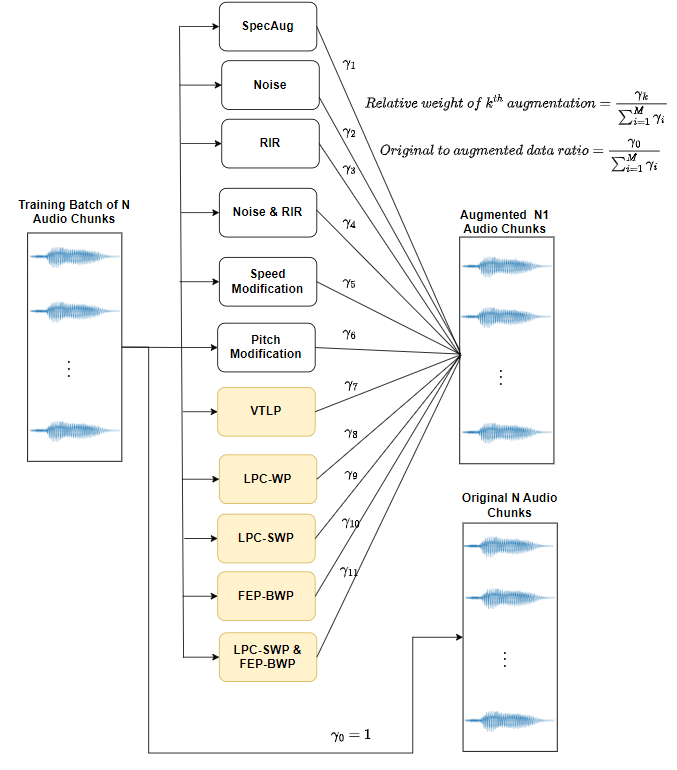}

\caption{Data augmentation methods for training speaker embedding extractor. Original data to augmentation ratio and relative weight of each augmentation method are also defined. 
Here, $\gamma_{i}$ is the weight of a particular augmentation method, $\gamma_{0}$ is the weight of original data, $N$ is the batch size, and $M$ is the number of augmentation methods. 
We present the augmentation methods and their weights ($\gamma_{i}$) in Table \ref{tab:res2}. While VTLP and LPC-WP were previously explored for children ASR in \cite{LPCAugment, kathania_aug}, we have investigated their impact on children's ASV tasks. Our work introduces a novel and more systematic algorithm for utilizing the subsequent method highlighted with shaded background, previously explored for children's ASR in \cite{me2022}, for children's ASV tasks. 
The second last method in the shaded background is novel and proposed in this paper. 
}
\label{fig:FIG1}

\end{figure}

\subsection{Contributions of This Work}

We present our proposed data augmentation pipeline in Fig. \ref{fig:FIG1}. 
First, we develop strong baselines including both application-agnostic data augmentations, namely, \emph{SpecAugment} \cite{specaugment}, \emph{Noise and RIR} \cite{musan,rir} and prosody-motivated data augmentations, namely, \emph{pitch modification (PM)} and \emph{speed modification (SM)} \cite{shahnawazuddin2020voice}. 

We then integrate vocal tract characteristics motivated data augmentation namely, vocal tract length perturbation (VTLP) \cite{Jaitly2013VocalTL} and linear prediction coding (LPC) phase warping (LPC-WP) \cite{LPCAugment}. 
Even if prior research already addresses the role of VTLP and LPC-WP data augmentation for children's speech processing, we identify two areas for detailed investigation. First, their prior investigation has focused exclusively on ASR tasks \cite{kathania_aug, LPCAugment}. Second, these techniques have not been benchmarked against the application-agnostic \cite{sb} and prosody-motivated augmentation techniques \cite{shahnawazuddin2020voice}. \textbf{Therefore, our first contribution is to address these shortcomings by conducting a thorough study of VTLP and LPC-WP in children's ASV}. 

\textbf{As our second contribution, we propose two new vocal-tract length motivated augmentation algorithms}. Both algorithms are motivated by the goal of minimizing the differences in vocal tract spectral parameters between children and adult speech \cite{ref5,ref6}. Unlike the previous related method \cite{me2022} which modifies the LPC spectra in the frequency domain, the proposed method implements formant-specific warping by directly modifying the linear prediction coefficients, providing improved control over the formant-specific warping and bandwidth modifications. 
We will collectively refer to the proposed data augmentation methods along with the VTLP and LPC-WP as \emph{ChildAugment}.

In addition to choosing the augmentation methods, another crucial practical consideration is determining the proportion in which the different augmentations (including the original, unaugmented data) should be mixed during training. Our study is set to answer this question -- how to balance the data optimally. \textbf{Hence, our third contribution involves empirically investigating the impact of different augmentation method proportions in the augmented data.} 


Apart from the speaker embedding extractor, the scoring methods play a vital role in determining the ASV system's accuracy and reliability. Different from non-parametric methods like cosine scoring, parametric approaches including PLDA (probabilistic linear discriminant analysis) \cite{kaldi-plda} and NPLDA (neural PLDA) \cite{nplda} involve a training phase and, therefore, offer adaptability. 
Unfortunately, in scenarios with limited availability of utterances from the target domain (e.g., children's speech case), these parametric methods may struggle to generalize. 
In light of this limitation, \textbf{our fourth contribution involves the investigation of a novel parametric ASV scoring approach designed to enable training with limited data.}

\section{Data Augmentation in Children ASR and ASV}



\emph{Data augmentation} involves applying predetermined transformations to training data, generating augmented samples that share the same content but exhibit distinct acoustic characteristics \cite{augsp}. This improves the model's generalizability by introducing more diverse range of instances during training \cite{aug_review}. Given the scarcity of children's speech corpora, data augmentation is frequently employed in children's ASR and ASV. In the following, we provide a concise overview of the available techniques.

\subsection{Children's ASR}
Data augmentation techniques for children's ASR can be broadly categorized into three groups. The first category includes application-agnostic methods, such as impulse responses, additive noise (with varied types of noises and SNRs) \cite{snrc}, and SpecAugment \cite{specaugment}. 
The second category includes prosody-motivated speed and pitch modification \cite{prosodyasr}. The third, more specialized category consists of techniques designed to address expected variations in vocal characteristics between adults and children \cite{Jaitly2013VocalTL, LPCAugment, me2022, f0modCASR}. 


\subsection{Children's ASV}
For children's ASV, only a few prosody-motivated in-domain data augmentation techniques have been explored. 
Specifically, speed and pitch modification for i-vector and x-vector based children's ASV have been studied in \cite{shah}. 
In addition, voice conversion based out-of-domain data augmentation has also been studied \cite{shah}. 
We focus entirely on in-domain data augmentation techniques as they do not necessitate any children's data as a prerequisite. Hence, they are applicable to zero-resource scenarios. We consider the following data augmentation techniques.

\textbf{\emph{SpecAugment}} modifies spectrograms by applying various time-warping, time-masking, and frequency masking \cite{specaugment}. Following \cite{sb}, we consider only time domain SpecAugment. { SpecAugment reduces the over-fitting \cite{specaugment}.} 
\textbf{\emph{Noise}}: The addition of background noise to the audio data, simulating real-world conditions \cite{musan}. { It improves the model in noisy environments.}
\textbf{\emph{RIR}}: Room impulse response-based augmentations that emulate different acoustic environments \cite{rir}. { RIR 
 augmentation improves the model in reverberant environments.}
\textbf{\emph{Noise and RIR}}: Combines Noise and RIR in this order in the time domain.
\textbf{\emph{SM}}: Speed modification, { explored earlier in children's ASV by \cite{shah}.}
\textbf{\emph{PM}}: Pitch modification, { explored earlier in children's ASV by \cite{shah}.}

We also explore existing data augmentation methods that were developed for achieving vocal tract alignment between children and adult speech for children's ASR systems.
\textbf{\emph{VTLP}}: Vocal tract length perturbation 
\cite{Jaitly2013VocalTL} {has been effective for ASV systems as well~\cite{vtlpforasv}.} 
\textbf{\emph{LPC-WP}} proposed in \cite{LPCAugment} for children's ASR, modifies the angle of each LPC filter's pole independently.

In addition, we explore two new data augmentation techniques as discussed in the next section. {Hyperparameters (warping factors) associated with all augmentation methods are presented in Table \ref{tab:warping} and are taken from the prior literature without changes.}

\begin{figure*}
\centering
\sbox{\measurebox}{%
  \begin{minipage}[b]{.42\textwidth}
  \subfloat
    []
    {\label{fig:0A}\includegraphics[width=6cm,height=7.5cm]{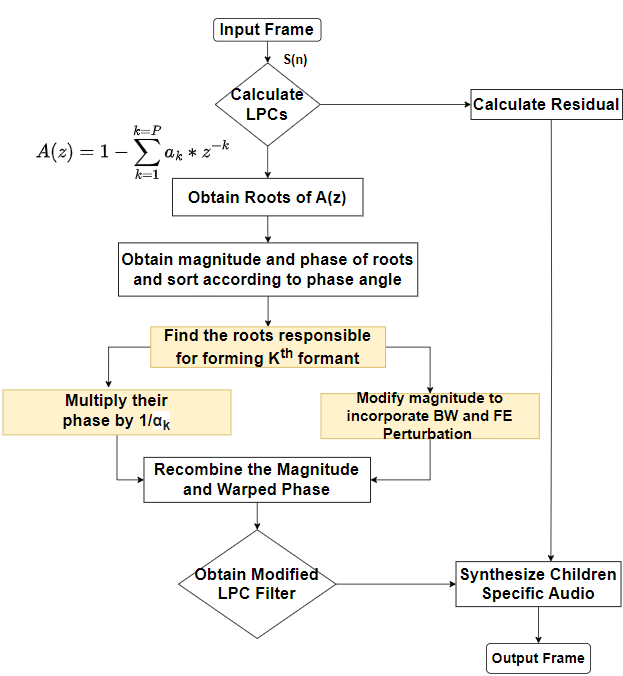}}
  \end{minipage}}
\usebox{\measurebox}\qquad
\begin{minipage}[b][\ht\measurebox][s]{.52\textwidth}
\centering
\subfloat
  []
  {\label{fig:0B}\includegraphics[width=4.5cm,height=4.5cm]{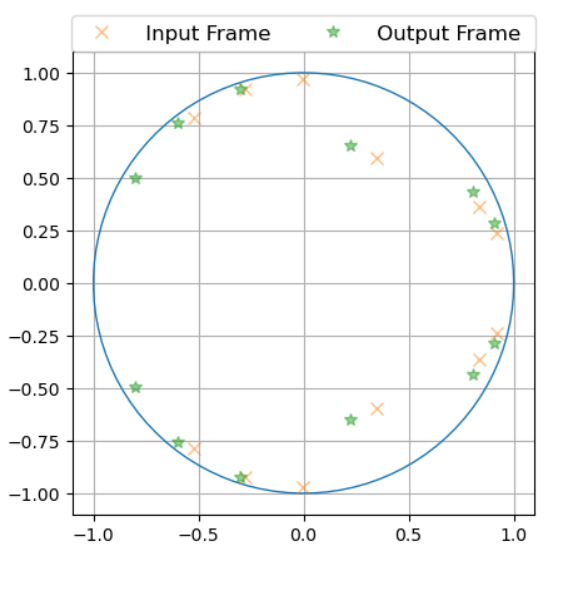}}

\vfill

\subfloat
  []
  {\label{fig:0C}\includegraphics[width=\textwidth,height=2cm]{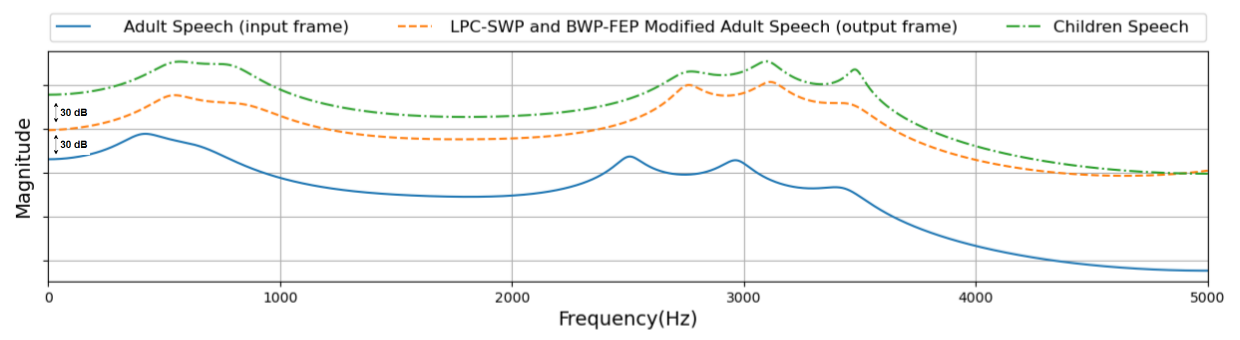}}
\end{minipage}
\caption{(a)~Acoustically motivated children-specific data augmentation pipeline. Shaded blocks indicate the key changes included for LPC-SWP and BWP-FEP modifications. $\alpha_k$ and $\beta_k$ are pole's phase warping and radius scaling factors for a set of poles responsible for forming $k$-th formant, respectively; (b)~Visualization of the original LPC roots for input frame, and corresponding LPC-SWP and BWP-FEP modified roots for output frames in the z-domain unit circle, for vowel /a/; and (c)~LPC spectra of an adult speech (input frame), LPC-SWP and BWP-FEP based modified adult spectra (output frames), along with the children spectra for a vowel /a/ computed in 25~ms window. Different LPC spectra are separated by 30~dB on the y-axis for better visualization. \label{fig:FIG0}}
\end{figure*}


\section{Vocal Tract Spectral Parameters Based Data Augmentation}\label{meth}
Previous studies \cite{ref5,ref6,jasa1} have highlighted the notable differences in vocal tract spectral parameters between adults and children. Primary factors that contribute to the observed mismatch between adult and children's speech can be summarized as follows:
\begin{enumerate} 
\item Formants of the children's LPC spectra exhibit non-uniform warping compared to those of the adult's spectra. Warping is defined as the transformation of the frequency axis, achieved by multiplying each frequency value with a constant called \emph{warping factor}. When the different frequency values are multiplied with different warping factors, then transformation is called \emph{non-uniform} warping.
\vspace{-0.0cm}
\item Relative formant amplitudes also differ substantially between adults and children. Specifically, the third and fourth formants in children's spectra tend to have higher amplitude than adult's spectra.
\vspace{-0.0cm}
\item The bandwidth of the formants in the children's LPC spectra tends to be {wider} than that in the adult's spectra \cite{formant_bandwidth}. This might be due to the incomplete development of vocal tract cavities responsible for speech production in children. The bandwidths of formants are influenced by multiple physical factors, including radiation, vocal tract wall compliance, viscosity, heat conduction, and glottal opening \cite{formant_bandwidth}. 
\end{enumerate}
\vspace{-0.0cm}
In a recent study \cite{me2022}, \emph{segmental warping perturbation} (SWP) and \emph{formant energy perturbation} (FEP) were proposed to address non-uniform warping and uneven energy distribution across formants. {The uniform warping algorithms, such as those proposed by \cite{f0modCASR}, incorporate uniform warping across all formants with the help of a single warping factor in a \emph{first-order all-pass filter} proposed in \cite{wlp}. On the other hand, the algorithm proposed by \cite{me2022} uses four different warping factors corresponding to the first four formants for incorporating formant-wise non-uniform warping. However, both SWP and FEP algorithms directly warp different parts of the LPC spectrum contour with different warping factors, leading to discontinuities in the LPC spectrum envelope.}

In this paper, we propose a novel non-uniform algorithm that incorporates segmental warping and formant energy perturbation directly in the LPC domain. {Our proposed method achieves formant-specific warping by modifying the linear predictor coefficients themselves. This provides greater control over the formant-specific warping process and ensures continuity in the modified LPC spectrum envelope.} Additionally, we introduce an algorithm for augmenting formant \emph{bandwidths}. {The proposed algorithms are illustrated in Figure \ref{fig:FIG0} and detailed below.}



\begin{table}[th]
\scriptsize
  \caption{Range of warping factors used in experiments. {The $\alpha$ and $\beta$ values, representing the range of warping factors for LPC-SWP and BWP-FEP, are derived from the formant data presented in \cite{ref6} and warping factors in \cite{me2022}  { and obtained sequentially $(i.e.\  \alpha_{1},\  then\  \alpha_{2})$ }. The range of $\alpha$ for the remaining methods is taken from the prior literature without changes.}} 
  \label{tab:warping}
  \centering
  \vspace{-0.05cm}
  \begin{tabular}{|c | p{3.3cm}|} 
  \hline 
  Experiments & \hspace{0.55cm} Warping Factors \\
  \hline\hline
  \multirow{1}{*}{SM\cite{shahnawazuddin2020voice}} & \hspace{0.85cm} $\alpha \in [0.9,1.1]$ \\ 
  \hline
  \multirow{1}{*}{PM\cite{shahnawazuddin2020voice}} & \hspace{0.85cm} $\alpha \in [0.9,1.1]$ \\
    \hline 
  \multirow{1}{*}{VTLP\cite{Jaitly2013VocalTL}} & \hspace{0.85cm} $\alpha \in [0.9,1.1]$ \\
    \hline  
 \multirow{1}{*}{LPC-WP \cite{LPCAugment}} & \hspace{0.85cm} $\alpha \in [0.7,1.3]$ \\
    \hline     
 \multirow{1}{*}{LPC-SWP {\textbf{(Proposed)}}} & $\alpha_1 \in [0.6, 0.85]$, $\alpha_2 \in [max(0.7,\alpha_1), 0.85]$, $\alpha_3\in [max(0.75,\alpha_2), 0.95]$, $\alpha_4 \in \hspace{0.65cm} [\max(0.85,\alpha_3), 1.0]$\\
    \hline  
 \multirow{1}{*}{BWP-FEP {\textbf{(Proposed)}}} & \hspace{0.1cm} $\beta_1 , \beta_2, \beta_3, \beta_4 \in [0.9, 1.1]$ \\
    \hline
 \end{tabular}
 \vspace{-0.0cm}
\end{table} 

\vspace{-0.0cm}
\subsection{LPC Segmental Warping Perturbation (LPC-SWP)}
\label{section:swp}
\vspace{-0.0cm}
We propose to augment the training dataset by modifying the different LPC coefficients of the adult's speech spectra non-uniformly, as follows:

\begin{enumerate} 
\item Obtain $p^\text{th}$ order LPC coefficients, $\boldsymbol{a}=(a_1,\dots,a_p)$, of windowed signal $s(n)$ using the auto-correlation method \cite{j.makhoul}.
\item Compute the residual $e(n)$ by passing the windowed signal $s(n)$ through the filter $A(z) = 1 - \sum_{k=1}^{k=p} a_{k} z ^{-k}$.
\vspace{-0.0cm}
\item Use a root finding algorithm on $A(z)$ to obtain complex conjugate roots $r_{i}$. The obtained roots can be either real or occur in conjugate pairs. 
The number of root pairs will be $p/2$ if $p$ is even. 
\item Find the complex conjugate root pairs responsible for forming the $k^{th}$ formant based on forward and backward slopes of neighboring angles and differences in their magnitude \cite{lpcf}. 
\vspace{-0.0cm}
\item Compute the magnitude and phase angle of each complex conjugate root responsible for forming the $k^{th}$ formant.
\vspace{-0.0cm}
\item Warp the phase angle of each complex conjugate pair roots forming the $k^{th}$ formant with the same warping factor $\nicefrac{1}{\alpha_k}$. For each frame, and for each formant, $\alpha_{k}$ is sampled independently from a uniform distribution range defined in Table \ref{tab:warping}. The modified roots are defined as: $r'_{i} = \lvert r_{i} \rvert e ^{j \frac{1}{\alpha_{k}} \angle{r_{i}}}$. Our motivation for warping different formants differently comes from previous studies on comparing children's and adults' speech characteristics \cite{ref5,ref6}. The range of warping factor $\alpha_k$ shown in  Table \ref{tab:warping} is obtained from the warping factor across the formants presented in \cite{ref5,ref6}.
\vspace{-0.0cm}
\item Repeat Step-6 for the first four formants. This ensures that each formant is warped differently and motivated by children's acoustics in \cite{ref5,ref6}. We restrict the warping to the first four formants since these formants are typically prominent in the LPC spectra of voiced speech. 
\vspace{-0.0cm}
\item Obtain the new filter  $A'(z)$ corresponding to the modified complex conjugate roots $r'_{i}$.
\vspace{-0.0cm}
\item Pass the residual obtained in Step-2 through the filter $\nicefrac{1}{A'(z)}$ to obtain the \emph{children}-\emph{specific} audio.
\end{enumerate}

\vspace{-0.0cm}
\subsection{Formant Bandwidth and Energy Perturbation (BWP-FEP)}
\label{section:fbp-fep}
\vspace{-0.0cm}
In order to modify the formant bandwidth, we use the known relationship \cite{bandwidth} between the pole radius $\lvert r_{i} \rvert$ and the corresponding 3-dB bandwidth $B$:
\vspace{-0.0cm}
\begin{align}
 \lvert r_{i} \rvert = e^{-\pi B T}.
 \label{eq:eq1}
\end{align}
After modifying the pole radius  $\lvert r_{i} \rvert$ responsible for forming the $K^{th}$ formant by factor $\beta_{k}$:
\vspace{-0.0cm}
\begin{align}
 \lvert r^{'}_{i} \rvert = \lvert \beta_{k} r_{i} \rvert= e^{-\pi B^{'} T}, \label{eq:eq2}
\end{align}
where $B^{'}$ is the modified 3-dB bandwidth and $T=1/f_s$ where $f_s$ is the sampling rate in Hz. For each frame, and for each formant, $\beta_{k}$ is sampled independently from a uniform distribution range defined in Table \ref{tab:warping}. It is important to note that the pole radius is also associated with formant energy. Therefore, the modification of formant bandwidth involves simultaneous modification of formant amplitude as well. One downside of this method is that the resulting filter is not guaranteed to be \emph{stable} (i.e. could have poles outside of the unit circle \cite{bandwidth}). To this end, our practical remedy to ensure a stable filter is to limit the pole radius to $1-\epsilon$, where $\epsilon = 0.02$.  

Figure \ref{fig:0B} showcases the representation of the original poles and corresponding modified poles (using LPC-SWP and BWP-FEP) on the unit circle in the $z$-domain. Figure \ref{fig:0C} displays various spectra, namely the LPC spectrum of adult speech, the modified LPC spectrum of adult speech employing acoustically motivated LPC-SWP and BWP-FE, and finally, the LPC spectrum of children's speech for vowel /a/. {For reproducibility, we provide the evaluation protocols and codes \href{https://github.com/vpspeech/ChildAugment}{here}.}

\begin{table}[t]
 \scriptsize
 \caption{CSLU kids Corpus (scripted): \# speakers and \# of utterances from different gender in the scripted set of CSLU kids corpus. We remove speakers with less than two utterances in the dataset to ensure at least one enrollment and one test audio per speaker, resulting in \emph{Cleaned} version as indicated in this table. We employ the Eval-good dataset to assess the ASV systems separately for boy and girl speakers. On the other hand, we utilize Dev-good to fine-tune the various scoring methods.}
 \vspace{-0.05cm}
 \label{tab:data}
 \centering
 \begin{tabular}{|c | c | c | c |c | c| c|} 
 \hline
 \multicolumn{1}{|c|}{} &\multicolumn{3}{c|}{Boys} & \multicolumn{3}{c|}{Girls}\\
\cline{2-7}
 Split & \# Spk & \# Utt  & {\# Hours} & \# Spk & \# Utt & {\# Hours}\\ 
 \hline\hline
Original Dataset & 605 & 39184 & {38.0} & 513 & 32815& {31.8} \\
\hline
Cleaned &  602 & 39167 & {37.9} & 511 & 32811 & {31.8}\\
\hline
With \textbf{good} tag & 602 & 25454 &  {22.6} &	511 &	22071 &	{19.7} \\ 
 \hline
 Eval-all & 542 & 35295 & {34.2} & 451 & 29082 & {28.2}\\ 
 \hline
 Dev-all & 60 & 3872 & {3.7} & 60	& 3729 & {3.6}\\ 
 \hline
Eval-good & 542 & 22960 & {20.4} & 451 & 19561 & {17.6}\\
 \hline
Dev-good & 60 & 2494 & {2.2} & 60 & 2510 & {2.2}\\ 
 \hline
 \end{tabular}
\hspace{-7.0 cm}
\end{table} 



\vspace{-0.0cm} 
\section{Experimental Setup}
\vspace{-0.0cm}
\subsection{Datasets}
\label{sec:data}
\vspace{-0.0cm}
{We consider two different speech corpora, namely, CSLU kids corpus \cite{cslu} and My Science Tutor (MyST) \cite{myst} for experimentation.} 
The first corpus, CSLU kids, includes both spontaneous and scripted recordings from around 1100 children. The collection involved a gender-balanced group of roughly 100 children per grade, spanning from kindergarten (5-6 years old) to grade 10 (15-16 years old). We utilize both spontaneous and scripted recordings from the CSLU kids corpus in our experimentation.

{The CSLU spontaneous subset contains recordings from 595 boys and 506 girls speakers. However, it is not straightforward to utilize them for the children's speaker verification task due to the presence of only one long audio recording (average duration 1 minute 39 secs) per speaker. Preparing the trials for ASV requires multiple utterances per speaker, which this dataset lacks. Hence, we utilized the open-source pre-trained Silero \emph{voice activity detection} (VAD) model \cite{Silero-VAD} to segment the longer audio recordings into an average of 30 audio chunks per speaker, each with an average duration of 1.4 seconds. 
The VAD-segmented CSLU spontaneous recordings are noisy and contain the teacher's noise in the background.} 

The scripted recordings contain isolated words and sentences from a total of 605 boys and 513 girls. We remove speakers with less than two utterances to ensure at least one enrollment and one test audio per speaker, resulting in \emph{Cleaned} version as indicated in Table \ref{tab:data}. The CSLU kids corpus does not contain a publicly-available standard evaluation protocol. We, therefore, designed customized, speaker-disjoint \emph{development} and \emph{evaluation} subsets presented as Dev-all and Eval-all (Table \ref{tab:data}). Whereas the former serves for fine-tuning the scoring methods, the latter consists of enrollment and test data for 542 boys and 451 girls, designated as the target speakers. 

The scripted speech recordings in the CSLU kids corpus are categorized into one of four groups: 1) \emph{Good}, where only the target word is spoken; 2) \emph{Maybe}, indicating the presence of the target word but with additional irrelevant content in the file; 3) \emph{Bad}, denoting that the target word is not spoken; and 4) \emph{Puff}, which is the same as ``Good" but with inhaling noise, characteristics of under-developed pronunciation skills in children \cite{jasa1,jasa2}. We use only the \emph{Good} recordings. 

The scripted subset of CSLU kids corpus consists of three types of utterances:
(i) Very short, single-word utterances with an average duration of 0.92 seconds 
(ii) Utterances containing multiple words, and
(iii) Utterances containing alpha-numeric words. 
We exclude the first type from our experiments. Additionally, we prepare three kinds of trial pairs for each gender:
(i) Comparing multiple-word sentences to multiple-word sentences (\textbf{S2S}),
(ii) Comparing alpha-numeric sentences against alpha-numeric sentences (\textbf{A2A}), and finally,
(iii) Comparing alpha-numeric sentences against multiple-word sentences (\textbf{A2S}). 

{Our second evaluation corpus, MyST, comprises 470 hours of English speech from 1371 students in grades 3-5 engaged in conversations with the tutor. MyST comprises three distinct speaker-disjoint subsets: training (1052 speakers), development (153 speakers), and testing (156 speakers). We utilize the test set of MyST for evaluation.} 

\begin{figure}[t]
\includegraphics[width=2.7in]{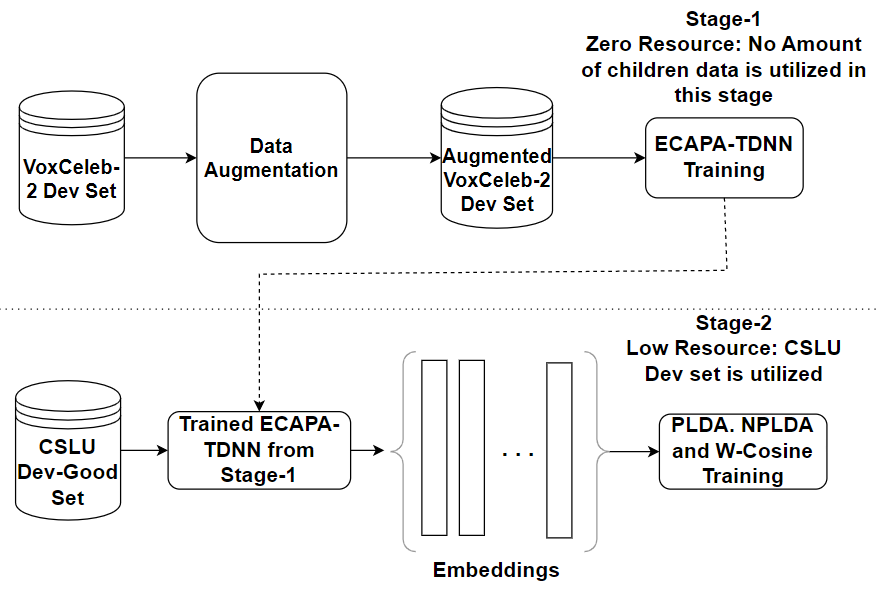}
\caption{{Block diagram representing the ECAP-TDNN training in zero-resource, and PLDA/NPLDA/W-Cosine training in low-resource scenarios.}}
\label{fig:FIG1.1}
\vspace{-0.0cm} 
\end{figure}

We train our speaker embedding extractors using the development set of VoxCeleb2 corpus \cite{vox2}, which consists of 5,994 speakers.
Additionally, we utilize babble noise from MUSAN \cite{musan} and room impulse response (RIR) \cite{rir} datasets for additive noise and RIR augmentations, respectively. {As shown in Fig. \ref{fig:FIG1.1}, no children's speech is used in training the embedding extractor --- a \emph{zero-resource} scenario.}

\vspace{-0.0cm} 
\subsection{Speaker Verification System}
\vspace{-0.0cm}
We conduct experiments using the SpeechBrain \cite{sb} toolkit with ECAPA-TDNN \cite{ecapa} based ASV system. The acoustic features consist of 80-dimensional mel-filterbank (Fbank) outputs extracted every 10 ms using a 25 ms window. The architectures of all ECAPA-TDNN models are the same as in \cite{ecapa}. They consist of frame-level convolutional layers with 1024 channels and 128 attention channels. The final, fully connected layer has 192 input nodes which is also the dimension of speaker embedding.

We train all the models for 12 epochs using the Adam optimizer \cite{adam}, with the learning rate ranging from ${1e}^{-3}$ to ${1e}^{-8}$. The learning rate scheduler follows a \emph{triangular} learning rate approach \cite{triang2} with cyclically varied learning rate between two pre-specified values. 
We use additive angular margin (AAM) softmax \cite{aam1,aam2} with a margin of 0.2 and softmax scaling of 30 for 4 cycles. Each cycle consists of 65,000 iterations. Additionally, a weight decay \cite{weightd} coefficient of ${2e}^{-5}$ is applied to all model weights. Similar to \cite{sb}, we randomly divide the dataset into a 90:10 ratio to create the training and validation sets. 

\begin{table*}[t]
\scriptsize
 \caption{Various models and weighting $(\gamma_{i})$ of particular augmentation methods when original to augmented data ratio is $1\colon3$ (define as $1\colon\sum\gamma_{i}$) and selected augmentation methods have equal weighting. If the weight is 0 then the particular augmentation is not considered in the corresponding model training. { The Baseline-$\nicefrac{x}{y}$ or Proposed-$\nicefrac{x}{y}$ has $x-$fold augmentation with equal weights of $y$ augmentation methods. As a result, $\gamma_{i} = \frac{\text{x}}{\text{y}}$, $x$ is set to 3 in this Table. Hence, $\gamma_{i}$ varies for different models based on the number of augmentation methods employed.} }
 \label{tab:res2}
\vspace{-0.0cm}
 \centering
 \begin{tabular}{|c | c | c | c |c |c |c|c|c|c|c|c|} 
 \hline
 \multicolumn{1}{|c|}{} &\multicolumn{4}{c|}{Application Agnostic }&\multicolumn{2}{c|}{Prosody}&\multicolumn{5}{c|}{Vocal Tract Characteristic Alignment} \\
 \cline{2-12}
 \multicolumn{1}{|c|}{} &\multicolumn{6}{c|}{\cite{sb,shah} }&\multicolumn{5}{c|}{ChildAugment} \\
\cline{2-12}
 Model & SpecAugment&  Noise & RIR & Noise + RIR &SM & PM & VTLP & WP & SWP & BWP-FEP & SWP+BWP-FEP\\ 
 \hline\hline
Baseline-\nicefrac{3}{1} &3&0&0&0&0&0&0&0&0&0&0 \\
\hline
Baseline-\nicefrac{3}{3} &$\nicefrac{3}{3}$&$\nicefrac{3}{3}$&$\nicefrac{3}{3}$&0&0&0&0&0&0&0&0	 \\ 
 \hline
Baseline-\nicefrac{3}{4} &$\nicefrac{3}{4}$&$\nicefrac{3}{4}$&$\nicefrac{3}{4}$&$\nicefrac{3}{4}$&0&0&0&0&0&0&0 \\
 \hline
 Baseline-\nicefrac{3}{5} &$\nicefrac{3}{5}$&$\nicefrac{3}{5}$&$\nicefrac{3}{5}$&$\nicefrac{3}{5}$&$\nicefrac{3}{5}$&0&0&0&0&0&0 \\
 \hline
Baseline-\nicefrac{3}{6} &$\nicefrac{3}{6}$&$\nicefrac{3}{6}$&$\nicefrac{3}{6}$&$\nicefrac{3}{6}$&$\nicefrac{3}{6}$&$\nicefrac{3}{6}$&0&0&0&0&0\\ 
 \hline
Proposed-\nicefrac{3}{7} &$\nicefrac{3}{7}$&$\nicefrac{3}{7}$&$\nicefrac{3}{7}$&$\nicefrac{3}{7}$&$\nicefrac{3}{7}$&$\nicefrac{3}{7}$&$\nicefrac{3}{7}$&0&0&0&0\\ 
 \hline
 Proposed-\nicefrac{3}{8} &$\nicefrac{3}{8}$&$\nicefrac{3}{8}$&$\nicefrac{3}{8}$&$\nicefrac{3}{8}$&$\nicefrac{3}{8}$&$\nicefrac{3}{8}$&$\nicefrac{3}{8}$&$\nicefrac{3}{8}$&0&0&0\\ 
 \hline
 Proposed-\nicefrac{3}{9} &$\nicefrac{3}{9}$&$\nicefrac{3}{9}$&$\nicefrac{3}{9}$&$\nicefrac{3}{9}$&$\nicefrac{3}{9}$&$\nicefrac{3}{9}$&$\nicefrac{3}{9}$&$\nicefrac{3}{9}$&$\nicefrac{3}{9}$&0&0\\ 
 \hline
 Proposed-\nicefrac{3}{10} &$\nicefrac{3}{10}$&$\nicefrac{3}{10}$&$\nicefrac{3}{10}$&$\nicefrac{3}{10}$&$\nicefrac{3}{10}$&$\nicefrac{3}{10}$&$\nicefrac{3}{10}$&$\nicefrac{3}{10}$&$\nicefrac{3}{10}$&$\nicefrac{3}{10}$&0 \\ 
 \hline
 Proposed-\nicefrac{3}{11} &$\nicefrac{3}{11}$&$\nicefrac{3}{11}$&$\nicefrac{3}{11}$&$\nicefrac{3}{11}$&$\nicefrac{3}{11}$&$\nicefrac{3}{11}$&$\nicefrac{3}{11}$&$\nicefrac{3}{11}$&$\nicefrac{3}{11}$&$\nicefrac{3}{11}$&$\nicefrac{3}{11}$ \\ 
 \hline

 \end{tabular}
  \vspace{-0.0cm} 
\end{table*} 
\begin{table}[t]
\scriptsize
 \caption{ Comparison of the proposed baseline systems (in terms of \%EER) with reported results in \cite{sb}. 
 {The Baseline-$\nicefrac{x}{5}$ has a $x$-fold augmentation with equal weights of \emph{five} augmentation methods. Note that these five methods are the same as those five methods outlined in ECAPA-TDNN training \cite{sb} and Table \ref{tab:res2}.} All the models were trained for 12 epochs from {scratch}, however the best checkpoint with the lowest validation EER was obtained after \{3,5,8,11\} epochs for Baseline-$\nicefrac{1}{5}$, Baseline-$\nicefrac{2}{5}$, Baseline-$\nicefrac{3}{5}$, and Baseline-$\nicefrac{4}{5}$. {We include the epochs corresponding to the best model in the table below.}}
 \label{tab:res1}
 \centering
 \vspace{-0.0cm}
 \begin{tabular}{|c | c | c | c|} 
 \hline
 Model & Vox1-O & Vox1-E & Vox1-H \\ 
 \hline\hline
\cite{sb} & 1.30 & 1.98 & 3.62 \\
\hline
Baseline-\nicefrac{1}{5} {(ep=3)} & 1.41 & 2.11 & 3.38  \\
\hline
Baseline-\nicefrac{2}{5} {(ep=5)} & 1.37 & 1.94 & 3.06 \\
\hline
Baseline-\nicefrac{3}{5} {(ep=8)} & 1.15 & 1.36 & 2.61  \\
\hline
Baseline-\nicefrac{4}{5} {(ep=11)} & 1.29 & 2.04 & 3.14  \\
\hline
 \end{tabular}
 \vspace{-0.0cm}
\end{table} 

\begin{table}[t]
\scriptsize
 \caption{Results (in terms of \%EER) for ASV models on scripted set of CSLU Boy and Girl Speakers using Cosine Scoring when original to augmented data ratio in training is $1 \colon 3$. { \textbf{S2S:} Comparing multiple-word sentences to multiple-word sentences, \textbf{A2A:} Comparing alpha-numeric sentences against alpha-numeric sentences, \textbf{A2S: } Comparing alpha-numeric sentences against multiple-word sentences, \textbf{Agg: } agglomerated EER on combined trials from S2S, A2A, and A2S. 
 }}
 \label{tab:res4}
 \vspace{-0.0 cm}
 \centering
 \begin{tabular}{|c | c | c | c | c |c |c |c|c|} 
 \cline{2-9}
 \multicolumn{1}{ c| }{} &\multicolumn{4}{c|}{Boy} &\multicolumn{4}{c|}{Girl} \\
\cline{1-9}

 Model & S2S & A2A & A2S& {Agg} & S2S & A2A & A2S & {Agg} \\ 
 \hline\hline
  Baseline-\nicefrac{3}{1}  & 18.29 & 11.61 & 16.79 & {17.98}& 22.57& 14.38& 21.66 &{22.13}\\ 
  Baseline-\nicefrac{3}{3} &16.83 & 10.27 &  15.33& {16.48}&18.95&  12.62& 17.47 &{18.61}\\
  Baseline-\nicefrac{3}{4} & 15.74 & 9.82 & 14.94 & {15.39}&17.42 & 11.25 & 16.68 & {17.08}\\  
  Baseline-\nicefrac{3}{5} & 15.43 & 9.54 & 14.16 & {15.18} &17.51 & 10.98 & 16.16&{17.10}\\  
  Baseline-\nicefrac{3}{6} & 14.88 & 9.16&  13.72& {14.54}&16.98& 10.62& 15.89&{16.65}\\ 
  Proposed-\nicefrac{3}{7} &  14.01 & 8.95 & 13.28& {13.71}&16.42 & 10.37 & 15.04&{16.05}\\
  Proposed-\nicefrac{3}{8} &  13.92& 9.08& 13.17&{13.64} &16.65&  10.41& 14.86&{15.28}\\ 
  Proposed-\nicefrac{3}{9} &  13.08 & 8.65 & 12.38& {12.88}&16.19& 10.02 & 14.38& {15.95}\\ 
  Proposed-\nicefrac{3}{10} & 13.22 &  8.37&  11.86 &{12.96} &15.48& 9.71 & 14.45 &{15.31}\\
  Proposed-\nicefrac{3}{11} & 12.90 & 8.10 & 11.98 & {12.59}&14.73& 9.54& 14.06&{14.42}\\
 \hline
 \end{tabular}
 \vspace{-0.0cm}
\end{table}


\begin{table}[t]
\scriptsize

 \caption{{ Ablation study with various data augmentation for assessing the impact of individual methods. \emph{No aug: } without augmentation, \emph{AA:} application agnostics,  \emph{Pros:} prosody motivated, \emph{ChAug:} ChildAugment, and \emph{Data} represents corpus used for training and augmentations. Evaluation protocols are the same as Table \ref{tab:res4}. Results with a model trained with MyST are shown for reference.}}
 \label{tab:ablation}
 \vspace{-0.0 cm}
 \centering
 \begin{tabular}{| c | c |  c| c | c | c |c |c |c|} 
 \cline{3-8}
 \multicolumn{2}{ c| }{} &\multicolumn{3}{c|}{{Boy}} &\multicolumn{3}{c|}{{Girl}} \\
\cline{1-8}

 {Methods}& {Data} &{S2S} & {A2A} & {A2S}& {S2S} & {A2A} & {A2S} \\ 
 \hline\hline
   {No aug} &  {Vox2} &  {18.86} & {11.78} & {17.52}& {22.38}& {15.01}&{21.89} \\ 
    {AA} &  {Vox2} & {15.74} & {9.82}& {14.94} & {17.42}& {11.25}& {16.68} \\
   {AA+Pros} &  {Vox2} & {14.88} & {9.16}&  {13.72}& {16.98}& {10.62}& {15.89}\\ 
  {ChAug} & {Vox2} &{15.22}&{9.34}&{14.09}&{16.88}&{10.49}&{16.08}\\
   {AA+Pros+VTLP} &  {Vox2} & {14.01}& {8.95}& {13.28}& {16.42}& {10.37}& {15.04}\\
   {AA+Pros+WP} &  {Vox2} & {14.11}& {8.90}& {13.41}& {16.39}& {10.48}& {15.16}\\
   {AA+Pros+SWP} &  {Vox2} & {13.98}& {8.81}& {13.20}& {16.44}& {10.31}& {14.88}\\
   {AA+Pros+BWP} &  {Vox2} & {14.17}& {9.11}& {13.31}& {16.58}& {10.64}& {15.01}\\
   {AA+Pros+ChAug} &  {Vox2} & {12.90}& {8.10}& {11.98}& {14.73}& {9.54}& {14.06}\\
  \hline \hline
  {AA} & {MyST} & {5.83} & {3.06} &  {4.93}& {6.38}& {3.44}& {5.29}\\
 \hline
 \end{tabular} 
 \vspace{-0.0cm} 
\end{table}

\begin{table}[t]
\scriptsize
 \caption{{Results (in terms of \%EER) on \textbf{MyST Corpus} for top baselines and proposed ASV models using cosine scoring. Baseline-\nicefrac{3}{5}$^{*}$ is trained entirely on the train set of MyST dataset for reference, while the remaining models are trained on VoxCeleb2.}}
 \label{tab:myst}
  \vspace{-0.0 cm}
 \centering
 \begin{tabular}{| c | c | c |c|} 
 \cline{2-4}
\multicolumn{1}{ c| }{}  &\multicolumn{1}{c|}{{Baseline-\nicefrac{3}{6}}} &\multicolumn{1}{c|}{{Proposed-\nicefrac{3}{11}}} &\multicolumn{1}{c|}{{Baseline-${\nicefrac{3}{5}}^{*}$}}\\
 \cline{2-4} \hline
  {MyST Test Set}&{31.15}&{28.93} & {12.83}\\
 \hline
 \end{tabular}
 \vspace{-0.0cm}
\end{table} 

\begin{table*}[t]
\scriptsize
 \caption{{Results (in terms of \%EER) on \textbf{CSLU Spontaneous Speech} for top baselines and proposed ASV models on CSLU boys and girls speakers from each grade using Cosine scoring. Baseline-\nicefrac{3}{5}$^{*}$ is trained entirely on the train set of MyST dataset for reference, while the remaining models are trained on VoxCelebb2.}}
 \label{tab:res1.1}
  \vspace{-0.0 cm}
 \centering
 \begin{tabular}{|c | c | c | c |c |c |c|c|c|c|c|c|c|c| c|} 
 \cline{3-13}
\multicolumn{2}{ c| }{} &\multicolumn{11}{c|}{Grades} \\
\cline{2-13}
\multicolumn{1}{ c| }{} &\multicolumn{1}{c|}{Model} &\multicolumn{1}{c|}{K}&\multicolumn{1}{c|}{1}&\multicolumn{1}{c|}{2}&\multicolumn{1}{c|}{3}&\multicolumn{1}{c|}{4}&\multicolumn{1}{c|}{5}&\multicolumn{1}{c|}{6}&\multicolumn{1}{c|}{7}&\multicolumn{1}{c|}{8}&\multicolumn{1}{c|}{9}&\multicolumn{1}{c|}{10}\\

\cline{2-13}
\hline
  \multirow{3}{*}{Boys}& {Baseline-\nicefrac{3}{6}} & {37.76}& {37.32} & {35.04} & {34.21} &  {31.13}& {29.24} & {26.82}& {21.96} &{17.74} & {16.02} & {16.68} \\
  &{Proposed-\nicefrac{3}{11}} & {35.68} & {36.03}& {33.29} & {32.96} & {29.32} & {27.68} & {25.03} & {20.18} & {16.25} & {14.87} &{15.22} \\
  \cline{2-13}
  &{Baseline-${\nicefrac{3}{5}}^{*}$}& {27.85} &{25.32}&{21.91}&{20.02}& {18.78}& {18.45}& {19.25}& {20.43} & {28.31}& {27.35}&{29.93}\\
 \hline
 \hline
 
  \multirow{3}{*}{Girls}&{Baseline-\nicefrac{3}{6}} & {40.98}& {40.79} & {40.64} & {37.14} & {36.38}& {31.96} & {28.15}& {28.71}& {23.60} & {23.19} & {23.36}\\
  &{Proposed-\nicefrac{3}{11}} & {38.78}& {39.12} & {38.65}& {35.22}& {34.09}& {29.87}& {26.53} & {26.88} & {21.55} & {21.06}&{21.37}\\
  \cline{2-13}
  &{Baseline-${\nicefrac{3}{5}}^{*}$}& {33.83} &{26.98}&{28.16}&{23.52}& {20.97}& {18.69}& {18.77}& {21.91} & {21.46}& {23.98}&{22.83}\\
 \hline
 \end{tabular}
 \vspace{-0.0cm}
\end{table*}




\begin{table*}[t]
\scriptsize
 \caption{{Same as \ref{tab:res1.1}, but for CSLU Scripted speech.}}
 \label{tab:res1.3}
  \vspace{-0.0 cm}
 \centering
 \begin{tabular}{|c | c | c | c |c |c |c|c|c|c|c|c|c|c| c|} 
 \cline{3-13}
\multicolumn{2}{ c| }{} &\multicolumn{11}{c|}{Grades} \\
\cline{2-13}
\multicolumn{1}{ c| }{} &\multicolumn{1}{c|}{Model} &\multicolumn{1}{c|}{K}&\multicolumn{1}{c|}{1}&\multicolumn{1}{c|}{2}&\multicolumn{1}{c|}{3}&\multicolumn{1}{c|}{4}&\multicolumn{1}{c|}{5}&\multicolumn{1}{c|}{6}&\multicolumn{1}{c|}{7}&\multicolumn{1}{c|}{8}&\multicolumn{1}{c|}{9}&\multicolumn{1}{c|}{10}\\

\cline{2-13}
\hline
   \multirow{3}{*}{Boys}&{Baseline-\nicefrac{3}{6}} & {27.16}& {25.62} & {25.48} & {24.92} &  {24.31}& {23.59} & {24.09}& {21.58}& {14.82} & {15.68} & {14.43}\\ 
   &{Proposed-\nicefrac{3}{11}} & {24.81} & {24.35}& {23.86} & {23.56} & {22.47} & {21.63} & {22.38} & {19.75} & {13.44} & {14.71} &{13.18} \\
   \cline{2-13}
   &{Baseline-${\nicefrac{3}{5}}^{*}$} & {14.28} & {14.09} & {15.39} &{14.10} & {14.57} & {14.13} & {15.96} & {14.61} & {13.55} & {13.74} & {13.28} \\
 \hline \hline
   \multirow{3}{*}{Girls}&{Baseline-\nicefrac{3}{6}} & {28.18}& {28.31} & {29.19}& {27.85} &  {25.96}& {25.67} & {28.03}& {24.94}& {21.75} & {19.92} & {19.09}\\ 
   &{Proposed-\nicefrac{3}{11}} & {26.90}& {26.23} & {27.31}& {25.75}& {24.11}& {23.41}& {26.14} & {23.08} & {19.82} & {18.24}&{17.68} \\
   \cline{2-13}
   &{Baseline-${\nicefrac{3}{5}}^{*}$} & {14.94} & {14.11} & {15.19} &{14.74} & {13.37} & {14.25} & {15.73} & {14.68} & {12.60} & {11.20} & {10.68} \\
 \hline
 \end{tabular}
 \vspace{-0.05cm} 
\end{table*}


\begin{table*}[t]
\scriptsize
 \caption{Results (in terms of \%EER and minimum decision cost function (mDCF) with target prior of 0.01 and unit costs of 1.0 for both miss and false alarm rates) for top baselines and proposed ASV models on scripted set of CSLU boys speakers using Cosine, Weighted Cosine, PLDA, and NPLDA scoring methods. { \textbf{S2S:} Comparing multiple-word sentences to multiple-word sentences, \textbf{A2A:} Comparing alpha-numeric sentences against alpha-numeric sentences, \textbf{A2S: } Comparing alpha-numeric sentences against multiple-word sentences.}}
 \label{tab:res5}
  \vspace{-0.0 cm}
 \centering
 \begin{tabular}{|c | c | c | c |c |c |c|c|c|c|c|c|c|c|c|c|c|} 
 \cline{4-15}
 \multicolumn{3}{ c| }{} &\multicolumn{4}{c|}{S2S} &\multicolumn{4}{c|}{A2A} &\multicolumn{4}{c|}{A2S} \\
\cline{2-15}

\multicolumn{1}{ c| }{} & \multicolumn{1}{ c| }{Model} &  \multicolumn{1}{ c| }{Result} &\multicolumn{1}{ c| }{Cosine} & \multicolumn{1}{ c| }{W-Cosine} & \multicolumn{1}{ c| }{PLDA} & \multicolumn{1}{ c| }{NPLDA}& \multicolumn{1}{ c| }{Cosine} & \multicolumn{1}{ c| }{W-Cosine} & \multicolumn{1}{ c| }{PLDA} & \multicolumn{1}{ c| }{NPLDA}& \multicolumn{1}{ c| }{Cosine} & \multicolumn{1}{ c| }{W-Cosine} & \multicolumn{1}{ c| }{PLDA} & \multicolumn{1}{ c| }{NPLDA}\\ 
 \cline{2-14}\hline
  \multirow{4}{*}{Boys}& \multirow{2}{*}{Baseline-\nicefrac{3}{6}} & EER &14.88& 13.97 & 8.43 & 8.78 &  9.16& 8.42 & 5.73& 6.06& 13.72 & 13.15 & 7.39&7.96\\ 
  & &  {mDCF}&  {0.0086} &  {0.0083} &  {0.0068} &  {0.0084} &  {0.0038} &  {0.0046} &  {0.0047}&  {0.0060}&  {0.0090} &   {0.0089} &  {0.0069} &  {0.0075} \\
  \cline{3-15}

  &\multirow{2}{*}{Proposed-\nicefrac{3}{11}} & EER &12.90& 12.23 & 7.38& 7.79& 8.10& 7.59& 5.01 & 5.28 & 11.98 & 11.47&6.77 &7.22\\
  & &  {mDCF}&  {0.0080} &  {0.0081} &  {0.066} &  {0.0081} &  {0.0032} &  {0.0030} &  {0.0049}&  {0.0058}&  {0.0085} &   {0.0083} &  {0.0067} &  {0.0072} \\
 \hline
 \hline

\multirow{4}{*}{Girls}& \multirow{2}{*}{Baseline-\nicefrac{3}{6}} & EER & 16.98 & 16.20 & 9.33 & 9.40 & 10.62 &  9.55 & 6.88&7.20& 15.89 & 14.75 & 9.21&9.58\\ 
& &  {mDCF}&  {0.0091} &  {0.0092} &  {0.0076} &  {0.0086} &  {0.0033} &  {0.0032} &  {0.0041}&  {0.0063}&  {0.0090} &   {0.0089} &  {0.0071} &  {0.0075} \\
  \cline{3-15}

&\multirow{2}{*}{Proposed-\nicefrac{3}{11}} & EER & 14.73& 13.85& 8.12& 8.55& 9.54&8.46& 6.12& 6.40 & 14.06& 13.28& 8.54&8.95 \\
& &  {mDCF}&  {0.0086} &  {0.0086} &  {0.0074} &  {0.0082} &  {0.0029} &  {0.0031} &  {0.0039}&  {0.0061}&  {0.0087} &   {0.0085} &  {0.0070} &  {0.0072} \\
 \hline

 \end{tabular}
 \vspace{-0.05cm} 
\end{table*}

\vspace{-0.0cm} 
\subsection{Scoring Methods}
\label{sec:cosine}
\vspace{-0.0cm}
We consider four different scoring strategies that vary in their complexity, as measured by the number of parameters. 
\emph{Cosine scoring} stands out as a training-free non-parametric method for speaker verification tasks computed as the angle between the length-normalized enrollment embedding $(\boldsymbol{\phi}_{e})$ and the test embedding $(\boldsymbol{\phi}_{t})$:
\vspace{-0.05cm}
\begin{align}
\tiny
S(\boldsymbol{\phi}_{e},\boldsymbol{\phi}_{t}) = \frac{\langle\boldsymbol{\phi}_{e},\boldsymbol{\phi}_{t}\rangle}{\Vert \boldsymbol{\phi}_{e} \Vert \Vert \boldsymbol{\phi}_{t} \Vert},
\label{eq:eq3}
\end{align}

where $\langle\cdot,\cdot\rangle$ and $\Vert\cdot \Vert$ denote the inner product between two vectors and the norm of a vector, respectively. \emph{Weighted cosine} extends Equation \eqref{eq:eq3} as: 
\vspace{-0.0cm}
\begin{align}
\scriptsize
S(\boldsymbol{\phi}_{e},\boldsymbol{\phi}_{t}) = \frac{\langle \boldsymbol{w} \odot \boldsymbol{\phi}_{e},\boldsymbol{w}\odot \boldsymbol{\phi}_{t}\rangle}{\Vert \boldsymbol{w}\odot\boldsymbol{\phi}_{e} \Vert \Vert \boldsymbol{w}\odot\boldsymbol{\phi}_{t} \Vert },
\label{eq:eq4}
\end{align}
where $\odot$ denotes element-wise multiplication, and $\boldsymbol{w}$ 
is a \emph{ weight vector} with the same dimension as the speaker embeddings. We obtain $\boldsymbol{w}$ by optimizing the loss given in Equation \eqref{eq:eq5}, along with the Adam optimizer \cite{adam}. The idea is to learn $\boldsymbol{w}$ that drives the target and non-target scores towards their ideal score values of +1 and -1, respectively. 
\vspace{-0.0cm}
\begin{align}
\scriptsize
 \mathcal{L} = \underset { \hspace{2 cm} \emph{\textbf{(Target\ Loss)}}}{\frac{1}{N_{\text{t}}}\sum_{k=1}^{k=N_{t}} (1 - \sum_{i=1}^{i=d} w_{i}  \boldsymbol{\phi}_{e}^{i,k}w_{i} \boldsymbol{\phi}_{t}^{i,k} ) } \nonumber \\  + \underset { \hspace{2.7 cm} \emph{\textbf{(Non-target\ Loss)}}} { \frac{1}{N_{\text{nt}}}\sum_{k=1}^{k=N_{nt}} (1 + \sum_{i=1}^{i=d} w_{i} \boldsymbol{\phi}_{e}^{i,k}  w_{i}  \boldsymbol{\phi}_{t}^{i,k} }) \nonumber \\ 
 \underset { \hspace{-4 cm} \emph{\textbf{(Regularization)}}} {\hspace{-4 cm}+\lambda  \sum_{i=1}^{i=d} {\lvert w_{i} \rvert}^{2} ,}\label{eq:eq5}
\end{align}

where, $w_i$ is the $i$-th dimension of $\boldsymbol{w},$ $N_{\mathrm{t}}$ and $N_{\mathrm{nt}}$ are the total number of target and non-target trials in a batch, $\boldsymbol{\phi}_{e},$ and $\boldsymbol{\phi}_{t}$ are $d$-dimensional enrollment and test embedding vectors. Finally, $\lambda$ is a regularization coefficient. 

Our \emph{probabilistic linear discriminant analysis} (PLDA) training configuration follows the standard Kaldi recipe \cite{povey2011kaldi} based on methods detailed in \cite{kaldi-plda}. To train the PLDA model, we utilize the Dev-good dataset which contains 120 speakers. { Consequently, this imposes a limitation on the maximum LDA dimension, which should be less than the number of classes \cite{bishop}. We set the LDA dimension to 119, which is the upper limit.} 

Our final scoring method is \emph{neural} PLDA (NPLDA) \cite{nplda}. Like for PLDA, we set the LDA dimension to 119 and employ the Dev-good dataset for training. 
We combine the male and female speakers from the Dev-good dataset to train gender-independent weighted cosine, PLDA, and Neural PLDA models.

\vspace{-0.0cm} 
\section{Results and Discussion}
\label{res}
\vspace{-0.0cm}
\subsection{Impact of Original to Augmented Data Ratio}
\vspace{-0.0cm}
First, we assess the performance of the ECAPA-TDNN model trained with varying original-to-augmented data ratios across the three standard VoxCeleb evaluation protocols \cite{vox1}, namely, VoxCeleb-$\{$O,E,H$\}$. The results shown in Table \ref{tab:res1} indicate that Baseline-$\nicefrac{3}{5}$ model, trained with original-to-augmented data ratio of $1 \colon 3$, outperforms the other models, including those reported by \cite{sb}.
Importantly, the training time is \emph{halved} over the standard training pipeline of \cite{sb} that uses $1 \colon 5$ augmentation ratio and converges slower. We hence fix the original to augmented data ratio of $1 \colon 3$ for the remaining experiments. 



\begin{figure}[t]
\includegraphics[width=2.5in]{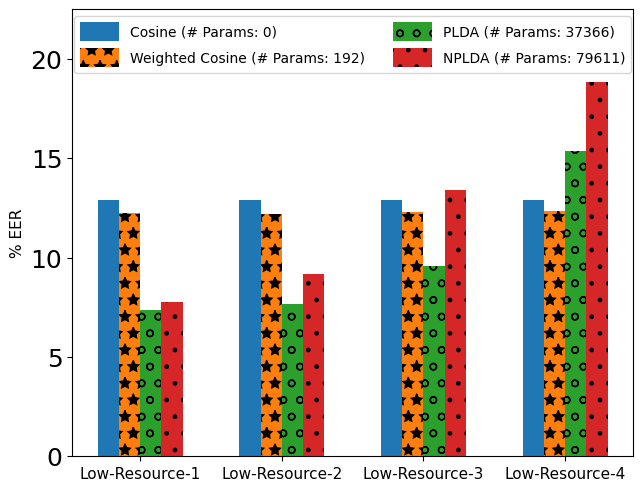}
\vspace{-0.0cm}
\caption{{Comparison of scoring methods in different low resource conditions. Results are presented for \emph{Proposed-$\nicefrac{3}{11}$} model on \textbf{S2S} evaluation set of CSLU boys speakers. (1) entire \emph{Dev-good} dataset (total 6.13 Hours), (2) 3 Hours from \emph{Dev-good} dataset, (3) 1.5 Hours from \emph{Dev-good} dataset, (4) 36 minutes from \emph{Dev-good} dataset. In all, 4 cases selected utterances were equally distributed across all speakers. The exact number of parameters in different scoring methods are also included. }}
\label{fig:FIG11}
\vspace{-0.05 cm} 
\end{figure}

\vspace{-0.0cm}
\subsection{Baseline Speaker Embedding Extractors}
\label{section:agno}
\vspace{-0.0cm}
Next, we present the {baseline} results with application-agnostic and prosody-motivated augmentations in Table \ref{tab:res4}. Baseline-$\nicefrac{3}{1}$, trained with $1 \colon 3$ original to augmented data ratio and SpecAugment only, yields high EER on all the three evaluation sets. Subsequently, incrementally adding the new augmentation method improves performance. Finally, Baseline-$\nicefrac{3}{6}$, trained with 6 augmentation methods, outperforms the other baselines on 5 out of 6 evaluation sets. This indicates that diversifying the training data with varied augmentation techniques is helpful, as one may expect. 

Additionally, both the baseline and the proposed models achieve the lowest EERs on the \emph{A2A} evaluation set for both genders, followed by the \emph{A2S} and \emph{S2S} sets. This could be due to longer-duration audios in the alpha-numeric sentences. { The lowest EERs on the \emph{A2A} evaluation set might also result from the limited vocabulary in alpha-numeric sentences. However, even in the \emph{A2S} task, which has a completely different vocabulary, the performance is better than the S2S task. Hence, the impact of longer audio duration can not be disregarded.} Further, across all the evaluation sets and for all the models, girl speakers consistently exhibit higher EER compared to boy speakers.

\vspace{-0.0cm}
\subsection{Impact of ChildAugment }
\label{section:childaug}
\vspace{-0.0cm}

Let us now address the {proposed models trained with} \emph{ChildAugment}. Tables \ref{tab:res2} and \ref{tab:res4} summarize the considered \emph{ChildAugment} variants and the obtained results, respectively. 
The Proposed-$\nicefrac{3}{7}$ model, which includes VTLP in addition to application-agnostic and prosody augmentations, outperforms the best baseline model on all the six trial sets. This indicates that modifying the vocal tract characteristics of adult speech indeed contributes to improving children's ASV performance. Further, the Proposed-$\nicefrac{3}{8}$ model, which additionally includes LPC-WP, exhibits marginal improvements over the Proposed-$\nicefrac{3}{7}$ model on three out of six evaluation sets.

Let us now focus on the proposed methods, LPC-SWP and BWP-FEP, discussed in Section \ref{meth}. The Proposed-$\nicefrac{3}{9}$ model, which includes LPC-SWP, outperforms both the Proposed-$\nicefrac{3}{7}$ and Proposed-$\nicefrac{3}{8}$ models by a considerable margin. Additionally, the Proposed-$\nicefrac{3}{10}$ model, which includes BWP-FEP, outperforms the Proposed-$\nicefrac{3}{9}$ model on four out of six evaluation sets. Lastly, the Proposed-$\nicefrac{3}{11}$ model, which combines LPC-SWP and BWP-FEP, achieves the lowest EER among all the models. {Specifically, the Proposed-$\nicefrac{3}{11}$ model, trained with ChildAugment in combination with application-agnostic and prosody-based data augmentation, yields the average relative improvements of 12.45\% (boys) and 11.96\% (girls) over Baseline-$\nicefrac{3}{6}$}. The improvements brought in by LPC-SWP and BWP-FEP indicate the importance of formant-wise warping and bandwidth modifications in aligning the vocal tract characteristics between adult and children speakers. 

{Thus far, we have considered ChildAugment in combination with other augmentation techniques. Now, let us focus on Table \ref{tab:ablation}, which contrasts ChildAugment against the other two broad categories of augmentations and \emph{without any augmentations.} As expected, the model trained without augmentation gives the highest EERs. Additionally, the model trained with ChildAugment alone outperforms the model trained with application-agnostic augmentations. Moreover, the model trained with ChildAugment alone also outperforms the model trained with a combination of application-agnostic and prosody augmentation in 2 out of 6 evaluation sets, underscoring its importance in comparison to established augmentation techniques. 

 {Further, we present fine-grained results for assessing the contribution of individual augmentation methods in ChildAugment. We observe that individual ChildAugment methods contribute almost equally when they are included along with application agnostic and prosody-based augmentations}. We also present the results for the model trained with the MyST \cite{myst} dataset for reference, yielding the lowest EERs, as might be expected.}

{Finally, we present the results for our best baseline and proposed models on MyST \cite{myst} test set in Table \ref{tab:myst}. The Proposed-\nicefrac{3}{11} model trained with ChildAugment outperforms the baseline model in this dataset as well. The models exhibit higher EER on MyST than the CSLU scripted set (Table \ref{tab:res4}). This might relate to the conversational nature of the MyST dataset and the presence of teacher cross-talk in the background. And, the third model trained on MyST train set data leads to a similar EER range as the CSLU scripted eval-set.}

\vspace{-0.0cm}
\subsection{Impact of Age}
\label{section:age-wise}
\vspace{-0.0cm}
{
Next, we present the age-wise results for the CSLU spontaneous and scripted corpus in Tables \ref{tab:res1.1} and \ref{tab:res1.3}, demonstrating a decrease in EERs with increasing age. The EER is highest for children in grades K and 1, and lowest for children in grades 9 and 10 for both genders. This suggests that as children age, their speech characteristics tend to approach those of adults, which can be attributed to the fact that the speaker embedding extractors are trained on VoxCeleb2, an adult speech dataset.  {Additionally, we observe a sharp decline in  EERs for boys in grades between 7 and 8. This could potentially be attributed to the onset of puberty for boys in grades 7 and 8.}} 

{For reference purposes, we also conduct age-wise analysis for an \emph{in-domain-oracle} model trained with the help of the largest children's corpus, MyST \cite{myst}. This enables us to assess the performance of our zero-resource models, trained with data augmentation, in comparison to an ECAPA-TDNN model (Model-MyST) exclusively trained on children's speech. As can be seen from Tables \ref{tab:res1.1} and \ref{tab:res1.3}, this in-domain-oracle model exhibits lower EER for age groups K to 7, while the Proposed-\nicefrac{3}{11} outperforms for the remaining age groups. This might be due to model bias for these age groups, as MyST contains speech from children between grades 5 to 8 only. Hence, augmentation methods provide uniform improvement across age groups and genders while the children-specific models are potentially sensitive to age and gender biases in the training data.}

\vspace{-0.0cm}
\subsection{Impact of Scoring Methods}
\label{section:scoring}
\vspace{-0.0cm}
Finally, we compare the four scoring methods in Table \ref{tab:res5}. Notably, PLDA scoring method outperforms the other scoring methods by a wide margin for both genders. While NPLDA outperforms both cosine scoring variants, it falls short of the PLDA results. This discrepancy may be attributed to the limited number of utterances available in the \emph{Dev-good} dataset relative to the large number of NPLDA parameters.

Despite falling short of the performance of PLDA and NPLDA, the weighted cosine backend outperforms standard cosine scoring, suggesting potential usefulness for low-resource scenarios where only a few development utterances are available. 
As the additional experiment in Fig. \ref{fig:FIG11} indicates, under the Low-Resource-4 condition, when only 10\% of the Dev-good dataset (36 minutes) is utilized in training, weighted cosine outperforms both PLDA and NPLDA. This advantage stems from the small number of required trainable parameters (specifically, 192 weights for our 192-dimensional speaker embeddings). 

\vspace{-0.0cm}
\section{Conclusions}
\vspace{-0.0cm}
We studied the impact of application-agnostic, prosody-motivated, and vocal tract characteristics driven data augmentation methods along with original to augmented data ratio for children's ASV. Our findings indicate that vocal tract characteristics driven data augmentation techniques, such as LPC-SWP and BWP-FEP, 
substantially enhanced the performance of children's ASV compared to models trained using application-agnostic and prosody-motivated data augmentation methods. Notably, these improvements were observed even when \emph{no children's audio} data was accessible for training the embedding extractor, thereby addressing the zero-resource scenario.  Additionally, we presented a new trainable \emph{weighted cosine scoring} with demonstrated potential over cosine, PLDA, and NPLDA scoring in \emph{extremely low resource} scenario. 

While our study provides a novel, practical approach to the role of data augmentation in children's ASV, we identify a number of limitations and future directions. 
First, in speaker verification, striking a balance between speaker embeddings with high between-speaker variability and low within-speaker variability is essential. Extreme modification of formant frequencies and bandwidths may lead to exaggerated intra-speaker variability which in turn may have a negative impact on accuracy. 
Second, we acknowledge the extensive range of training possibilities, where employing an ECAPA-TDNN training pipeline with $M$ augmentation methods could result in $2^M$ potential combinations and experiments. Due to the exponential growth with $M$, it becomes impractical to thoroughly explore \emph{all possible combinations of data augmentations.} 
In this study, we resorted to a specific incremental heuristic to choose specific data augmentation combinations; in future, we would like to explore an alternative approach using techniques such as reinforcement learning to determine the optimal combination, weighting of various augmentation methods, {and hyperparameters (such as warping factors and signal-to-noise ratio)} as an integral part of the model training process. {Additionally, given that our experiments were carried out on English datasets, it would be important to address the impact of language in our future work. Finally, while we focused solely on data augmentation, it would be also relevant to address the role of acoustic features and choice of the neural speaker embedding extractor in children's ASV tasks. We hope our work to serve as a viable starting point for such further explorations.}
\vspace{-0.4cm}
\section*{Author Declaration}
\vspace{-0.0cm}
The authors declare no conflict of interest.

\vspace{-0.4cm}
\section*{Data Availability}
\vspace{-0.0cm}
Data sharing is not applicable to this article as no new data were created or analyzed in this study.

\end{document}